\documentclass[12pt]{article}
\usepackage{epsfig}
\newcommand{\mysection}{\setcounter{equation}{0}\section}

\def\beq{\begin{equation}}
\def\eeq{\end{equation}}
\def\beqa{\begin{eqnarray}}
\def\eeqa{\end{eqnarray}}
 
\newlength{\dinwidth} \newlength{\dinmargin}
\begin{document}
 
\begin{center}
{\Large \bf 
Higgs Production via $b {\bar b} \rightarrow H$ in Hadron Colliders}
\end{center}
\vspace{2mm}
\begin{center}
{\large Nikolaos Kidonakis\footnote{Presented at DIS 2008, London, England, 
April 7-11, 2008}}\\
\vspace{2mm}
{\it Kennesaw State University, Physics \#1202\\
1000 Chastain Rd., Kennesaw, GA 30144-5591}
\end{center}

\begin{abstract}
I present NNNLO collinear and soft gluon corrections for Higgs production
via the process $b{\bar b} \rightarrow H$. I show that the collinear 
corrections dominate and contribute large enhancements to the cross section 
at both the Tevatron and the LHC.
\end{abstract}

\thispagestyle{empty}  \setcounter{page}{1}

\mysection{Introduction}

The discovery of the Higgs boson is a goal of paramount importance at 
hadron colliders. 
The search for the Higgs continues at the Tevatron and 
will soon start at the LHC.
The main Standard Model production channel at both colliders  
is $gg \rightarrow H$. The channel $b{\bar b} \rightarrow H$ 
is relatively small in the Standard Model but is numerically 
important in the MSSM at high $\tan \beta$.

The process $b{\bar b} \rightarrow H$ has very simple color structure 
and kinematics, essentially the same as for the Drell-Yan process \cite{url}. 
The complete QCD corrections are known to NNLO \cite{HK}.
The collinear \cite{NKHiggs} and soft gluon \cite{NKHiggs,NKNNNLO,VR} 
corrections can be calculated to higher orders.
Here we present results for these corrections through NNNLO.
It is shown at NLO and NNLO that the soft-gluon approximation is inadequate;
purely collinear terms must also be added to obtain a good approximation 
to the full QCD corrections.
We calculate the complete soft corrections 
and the leading and some subleading purely collinear terms at NNNLO 
\cite{NKHiggs}, and show that these higher-order corrections 
provide additional significant enhancements to the cross section at 
both the Tevatron and the LHC. 

\mysection{NNNLO collinear and soft corrections}

We first provide the analytical form of the soft and collinear corrections. 
We define $s=(p_b+p_{\bar b})^2$ and $z=m_H^2/s$, with $m_H$ the Higgs mass.  
As we approach threshold, $z \rightarrow 1$. 
The $n$-th order soft and collinear corrections in the partonic cross section 
are of the form 
\beqa
{\hat \sigma}^{(n)}_{S+C}(z)&=&\sum_{k=0}^{2n-1} S_k^{(n)}
\left[\frac{\ln^k(1-z)}{1-z}\right]_+ +\sum_{k=0}^{2n-1} C_k^{(n)}
\ln^k(1-z) 
\nonumber
\eeqa
where the first and second sums are, respectively, the soft corrections 
and the purely collinear corrections.

The NLO soft and collinear gluon corrections are given explicitly by
\beqa
{\hat\sigma}^{(1)}_{S+C}(z)
&=&F^B
\frac{\alpha_s(\mu_R^2)}{\pi} \left\{
4 C_F \left[\frac{\ln(1-z)}{1-z}\right]_+
-2 C_F \ln\left(\frac{\mu_F^2}{m_H^2}\right) \left[\frac{1}{1-z}\right]_+
\right.
\nonumber \\ && \hspace{20mm}\left. 
{}-4 C_F \ln(1-z) +2 C_F \ln\left(\frac{\mu_F^2}{m_H^2}\right)+2 C_F\right\}
\nonumber
\label{NLO}
\eeqa
where $F_B$ is the Born term, $C_F=4/3$, and $\mu_F$ is the factorization scale.
The NNLO soft and collinear gluon corrections are 
\beqa
{\hat\sigma}^{(2)}_{S+C}(z)
&=&F^B \frac{\alpha_s^2(\mu_R^2)}{\pi^2}
\left\{8 C_F^2
\left[\frac{\ln^3(1-z)}{1-z}\right]_+
+\cdots
-8 C_F^2 \ln^3(1-z)
+\cdots
\right\}
\label{NNLO}
\nonumber
\eeqa
and the NNNLO soft and collinear gluon corrections are
\beqa
{\hat\sigma}^{(3)}_{S+C}(z)&=&F^B \frac{\alpha_s^3(\mu_R^2)}{\pi^3}
\left\{8 C_F^3
\left[\frac{\ln^5(1-z)}{1-z}\right]_+
+\cdots
-8 C_F^3 \ln^5(1-z) 
+\cdots \right\}
\nonumber
\label{NNNLO}
\eeqa
where for brevity we only show the leading terms.

We now present results for the cross sections for 
$b {\bar b} \rightarrow H$ at the Tevatron, with $\sqrt{S}=1.96$ TeV,
and the LHC, with $\sqrt{S}=14$ TeV. 
We use the bottom quark parton distribution functions (pdf) from
the MRST2006 NNLO set of parton densities \cite{MRST2006}.
We are particularly interested in the relative size of the
higher-order contributions to the total cross section.
In the results below we denote the factorization and renormalization scales
by $\mu$.

\begin{center}
\epsfig{file=Kbbhtevnloplot.eps,height=0.39\textwidth,angle=0}
\hspace{1mm}
\epsfig{file=Kbbhlhcnloplot.eps,height=0.39\textwidth,angle=0}
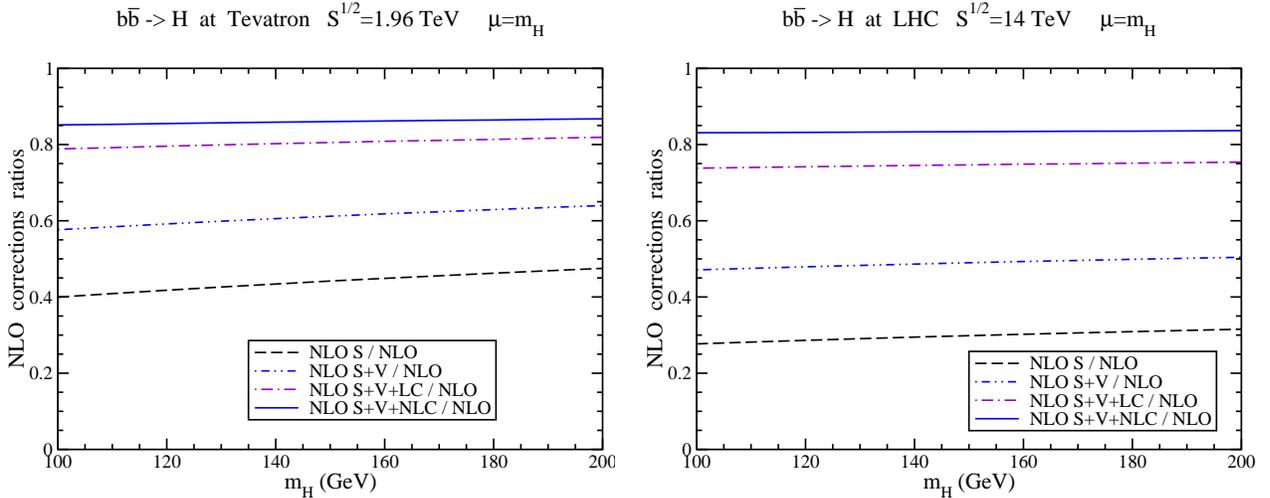
Figure 1: The NLO ratios for $b {\bar b} \rightarrow H$ 
at the Tevatron (left) and the LHC (right).
\end{center}

In Figure 1 we show the contribution of various terms
to the complete NLO corrections for Higgs production at the
Tevatron (left) and the LHC (right), setting $\mu=m_H$.
In this figure NLO denotes the ${\cal O}(\alpha_s)$ corrections only 
(i.e. without the Born term). The curve
marked NLO S / NLO denotes the fractional contribution of the NLO soft
(S) corrections to the total NLO corrections. We see that this contribution
does not surpass 48\% at the Tevatron and 32\% at the LHC 
and thus the soft-gluon approximation is by itself
inadequate. Adding on the virtual terms, the soft plus
virtual (S+V) corrections still do not provide a good approximation 
to the full corrections. Further adding collinear corrections
substantially improves the situation. When leading collinear (LC) logarithms 
are included, the resulting
S+V+LC approximation accounts for about 80\% of the total NLO corrections 
at the Tevatron and 75\% at the LHC. 
If we further add the next-to-leading collinear (NLC) terms the approximation
(S+V+NLC) gets even better, reaching around 85\% of the total corrections, at 
both the Tevatron and the LHC. 
Clearly the inclusion of collinear terms greatly improves the approximation
in both cases, and in particular it is
important to include the NLC terms.

\begin{center}
\epsfig{file=Kbbhtevnnloplot.eps, 
        height=0.39\textwidth,angle=0}
\hspace{1mm}
\epsfig{file=Kbbhlhcnnloplot.eps, 
        height=0.39\textwidth,angle=0}
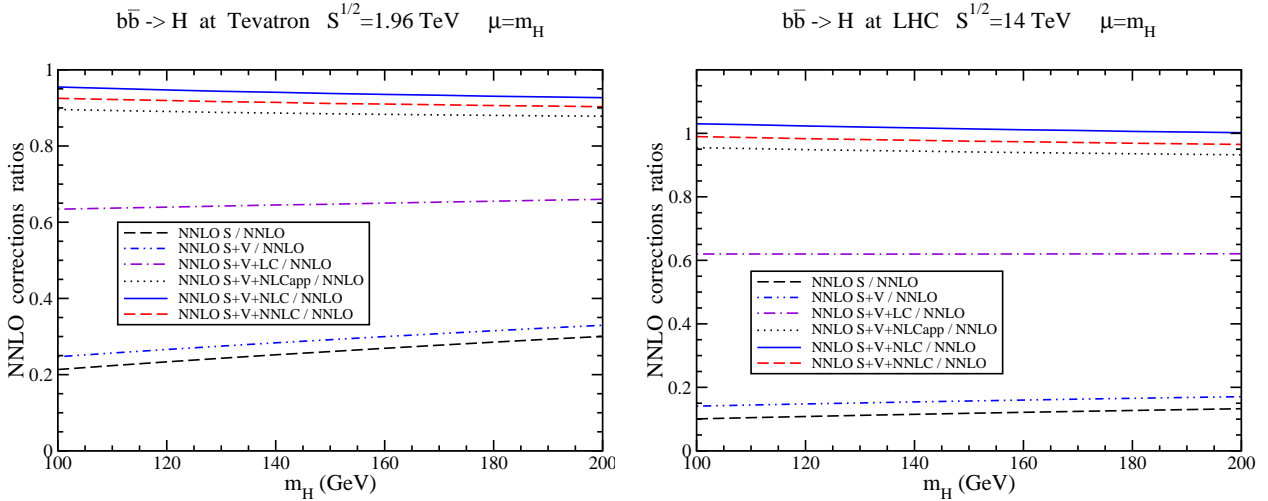
Figure 2: The NNLO ratios for $b {\bar b} \rightarrow H$ at 
the Tevatron (left) and the LHC (right).
\end{center}

In Figure 2 we show the contribution of various terms
to the complete NNLO corrections for $b {\bar b} \rightarrow H$ 
at the Tevatron (left) and the LHC (right), with $\mu=m_H$.
In this figure NNLO denotes the ${\cal O}(\alpha_s^2)$
corrections only (i.e. without the Born term and NLO corrections).
The curve NNLO S / NNLO denotes the fractional contribution of the
NNLO soft corrections to the total NNLO corrections. At the
Tevatron and the LHC the soft contribution is rather small, in fact even
smaller than at NLO. The same is true of the S+V contribution. 
The additional inclusion of the leading collinear logarithms accounts for
about 60\% of the total NNLO corrections at both the Tevatron and the LHC, 
which is better but still not satisfactory.
However, the further inclusion of the next-to-leading collinear logarithms 
improves the approximation significantly. The effect of the NLC terms is 
much more important at NNLO than at NLO, and thus the NLC terms are needed 
to achieve a good approximation.
We also plot a curve (S+V+NNLC) that in addition includes the 
next-to-next-to-leading collinear terms (NNLC).
We see that the NNLC terms alone do not make a large
contribution, and that the S+V+NNLC results approximate the exact NNLO
corrections very well.

\begin{center}
\epsfig{file=Kbbhtevplot.eps, 
        height=0.39\textwidth,angle=0}
\hspace{1mm}
\epsfig{file=Kbbhlhcplot.eps, 
        height=0.39\textwidth,angle=0}
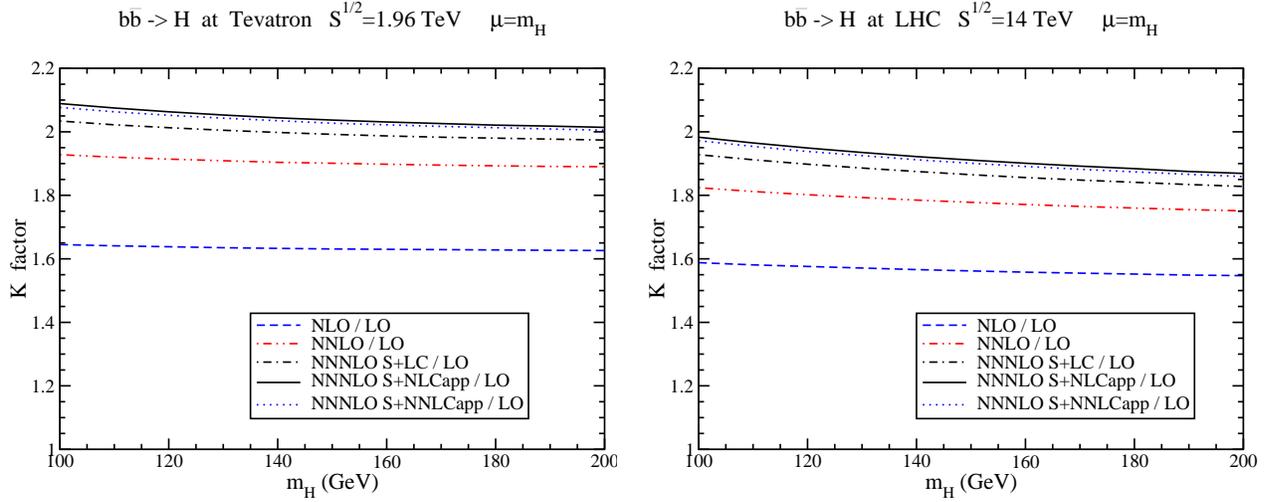
Figure 3: The $K$ factors for $b {\bar b} \rightarrow H$ at 
the Tevatron (left) and the LHC (right).
\end{center}

Figure 3 shows the $K$ factors at the Tevatron (left)
and the LHC (right), with $\mu=m_H$.
Here N$^k$LO cross section means
the Born term plus all the corrections through ${\cal O}(\alpha_s^k)$.
The NLO / LO curve shows that the complete NLO corrections increase the LO 
result by around 60\% at both the Tevatron and the LHC. Inclusion
of the complete NNLO corrections futher significantly increases the 
cross section: the NNLO $K$ factor is around 1.9 at the Tevatron and
1.8 at the LHC. Finally, we include the complete soft and approximate 
collinear corrections at NNNLO, which provide further significant 
enhancements. We plot one curve with the soft and leading collinear (S+LC)
terms, another curve with the soft and approximate NLC
(S+NLCapp) terms, and a third with the soft and approximate NNLC 
(S+NNLCapp) terms.
We note that the difference between the S+LC and S+NLCapp curves is small, 
and between the S+NLCapp and S+NNLCapp curves it is much smaller.
The NNNLO S+NNLCapp $K$ factor is between 2.00 and 2.08 at the
Tevatron and between 1.86 and 1.97 at the LHC for Higgs masses between
100 and 200 GeV. 
The conclusions from the study of the soft and collinear terms at NLO and NNLO 
at both the Tevatron and the LHC gives us confidence that the 
NNNLO S+NNLCapp curves
provide a good approximation of the complete NNNLO cross section.

\begin{center}
\epsfig{file=bbhtevplot.eps, 
        height=0.39\textwidth,angle=0}
\hspace{1mm}
\epsfig{file=bbhlhcplot.eps, 
        height=0.39\textwidth,angle=0}
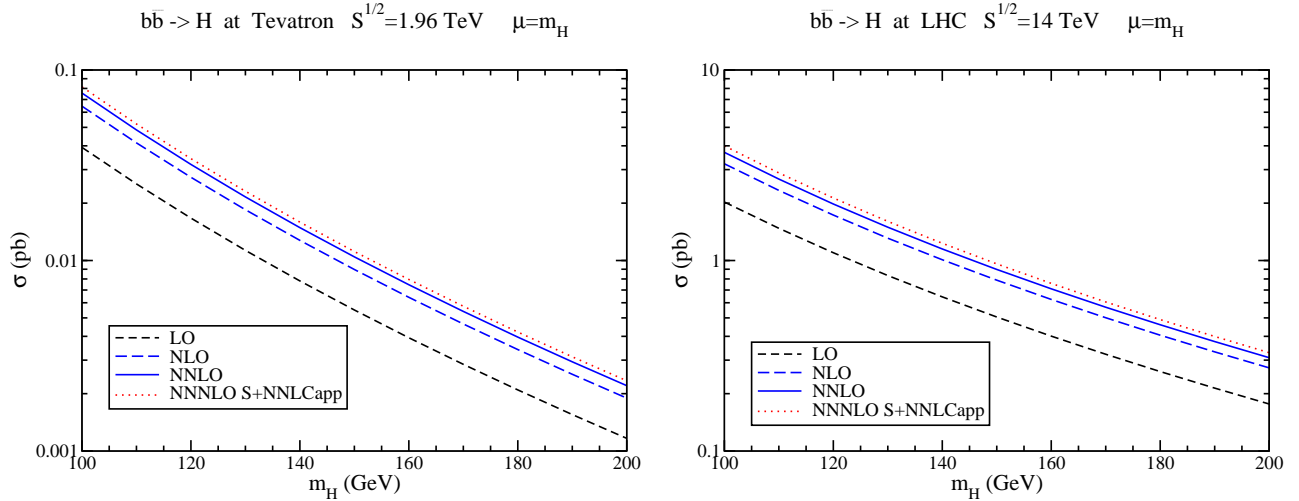
Figure 4: The cross section for $b {\bar b} \rightarrow H$ at 
the Tevatron (left) and the LHC (right).
\end{center}

In Figure 4 we plot the Standard Model cross sections for 
$b {\bar b} \rightarrow H$ at
the Tevatron (left) and the LHC (right). We show LO,
NLO, NNLO, and NNNLO S+NNLCapp results for $\mu=m_H$. 

\mysection*{Acknowledgements}

This work was supported by the 
National Science Foundation under Grant No. PHY 0555372.

\end{document}